**Does Antibiotic Resistance Evolve in Hospitals?**

Anna Seigal[1], Portia Mira[2], Bernd Sturmfels[1], Miriam Barlow[2]

[1]University of California, Berkeley, Department of Mathematics
[2]University of California, Merced, School of Natural Sciences

**Abstract:**
Nosocomial outbreaks of bacteria are well-documented. Based on these incidents, and the heavy usage of antibiotics in hospitals, it has been assumed that antibiotic resistance evolves in hospital environments. To test this assumption, we studied resistance phenotypes of bacteria collected from patient isolates at a community hospital over a 2.5-year period. A graphical model analysis shows no association between resistance and patient information other than time of arrival. This allows us to focus on time course data.

We introduce a Hospital Transmission Model, based on negative binomial delay. Our main contribution is a statistical hypothesis test called the Nosocomial Evolution of Resistance Detector (NERD). It calculates the significance of resistance trends occurring in a hospital. It can inform hospital staff about the effects of various practices and interventions, can help detect clonal outbreaks, and is available as an R-package.

We applied the NERD method to each of the 16 antibiotics in the study via 16 hypothesis tests. For 13 of the antibiotics, we found that the hospital environment had no significant effect upon the evolution of resistance; the hospital is merely a piece of the larger picture. The p-values obtained for the other three antibiotics (Cefepime, Ceftazidime and Gentamicin) indicate that particular care should be taken in hospital practices with these antibiotics. One of the three, Ceftazidime, was significant after accounting for multiple hypotheses, indicating a trend of decreased resistance for this drug.

## 1. Introduction

Antibiotic resistance is a global problem that results from selective pressures imposed by antibiotic consumption on an industrial scale (Rodriguez-Rojas, Rodriguez-Beltran et al. 2013). Most antibiotic consumption occurs in agricultural, clinical, and outpatient settings. Immigration of resistant strains throughout the world has made the emergence of resistance to antibiotics a global problem (Goossens 2009, Forslund, Sunagawa et al. 2013). However, regional differences in the first appearance of resistant genotypes and their subsequent frequencies indicate the importance of local factors (Kahlmeter, Ahman et al. 2015). It is unclear whether global population dynamics, regional factors, or immediate proximity to antibiotics determines the frequencies of specific resistance phenotypes in a defined location, such as a hospital. Any insight into this problem could have major effects on public health policy.

The success of a hospital-centered effort to reduce antibiotic resistance depends on whether antibiotic resistance is actually evolving in hospitals. Anecdotal evidence supports this assumption. For example, in 2011 the U.S. National Institutes of Health Clinical Center experienced an outbreak of carbapenam-resistant *K. pneumonia* that affected 18 patients, 11 of whom died (Snitkin, Zelazny et al. 2012). This scenario illustrates that a specific bacterial strain from a single patient can become endemic within hospitals, and likely evolve within the hospital environment.

Typically when evolution of antibiotic resistance is studied within a hospital, it is either with respect to clonal outbreaks (Lopez-Camacho, Gomez-Gil et al. 2014), or to evolution in individual patients with long-term infections (Zhao, Schloss et al. 2012). While important, these studies do not address the majority of transmission events in hospitals. For less virulent strains, the associated transmission of antibiotic resistance goes undetected. Such transmission is harder to track and deserves attention.

Many hospitals have attempted to reduce the frequencies of resistant isolates, with mixed success. Cycling antibiotics in individual spinal cord patients has shown promising results for preventing the emergence of multi-drug resistant urinary tract infections (Poirier, Dinh et al. 2015). In a review article (Brown and Nathwani 2005) analyzing the efficacy of cycling, the results showed that decreasing consumption of aminoglycosides in hospitals tends to reduce resistance to them. However, for β-lactam antibiotics there is no clear trend of reduced resistance in response to reduced consumption. This may be due to many factors, including the choice of antibiotics, the duration of therapies, and outside factors such as immigration into the hospital from the surrounding community.

There have also been attempts to ameliorate the resistance problem at larger scales than a single hospital. These too have delivered mixed results. Despite a nationwide effort to reduce β-lactam prescriptions in Turkey, β-lactam resistance increased, except for carbapenam resistance in *Pseudomonas* and *Acinetobacter*. The frequency of Methicillin Resistant *Staphylococcus aureus* (MRSA) also decreased (Altunsoy, Aypak et al. 2011). Efforts in agricultural settings have seemed promising. In Denmark, an agricultural ban of growth promoting antibiotics resulted in a significant decrease in the frequency of ampicillin, nalidixic acid, sulfonamide, tetracycline, erythromycin, and streptomycin resistant bacteria in food animals (World Health Organization 2002). The ban also resulted in a decrease in Vancomycin Resistant Enterococci (VRE) in both animal and human populations (Casewell, Friis et al. 2003).

We seek to answer the question of whether evolution of antibiotic resistance occurs in a hospital. By "evolution", we mean that the historical resistance phenotypes present in the hospital have a causative effect on the resistance phenotypes that are subsequently present in the hospital. The alternative is that antibiotic resistance is due to immigration into the hospital from the surrounding area. If resistance is due to immigration, then the distribution of resistance phenotypes in the hospital will be the proportions present in the wider population. On the other hand, if resistance evolves in the hospital, the distribution will deviate from such proportions. The relevant dynamics are discussed in Section 4.

Transmission of resistance in the hospital can occur via horizontal gene transfer between strains of bacteria that are either infectious or commensal (Machado, Coque et al. 2013, Hamidian and Hall 2014). It can also occur via transfer of bacteria between individuals, both patients and health care workers, through either direct or indirect contact.

Various mathematical models have been created to test the influences of different elements in the hospital (Austin and Anderson 1999, Lipsitch, Bergstrom et al. 2000, Gandon, Day et al. 2016). Testing resistance trends using such models requires a hospital to gather extra data. We develop a simpler model, based on data that hospitals typically gather, for determining whether transmission of resistance occurs in the hospital.

We conducted a case study with the Dignity Health Mercy Medical Center, a small community hospital in the Central Valley of California. Collaboration with a hospital was critical for this study because HIPAA patient privacy laws prevent patient data from becoming publicly available. We had access to de-identified patient data associated with ESBL strains for a 2.5 year interval. We developed a mathematical model and hypothesis test for the emergence of resistance. This is available as a statistical software package in R. We analyzed temporal data of the resistance phenotypes of patient isolates. The unit of analysis in this study is the resistance phenotypes of patient isolates. Although there may be multiple resistance genes that confer each phenotype, aggregating based on phenotype allows us to see that one of the resistance genes was present (Woodford, Ward et al. 2004, Appelbaum 2006, Bush 2010, Redgrave, Sutton et al. 2014). While it would be ideal to have genomic sequence data to accompany resistance data, those data are not currently available to us, nor are they commonly available to most hospitals, particularly those in developing nations. Additionally, despite the widespread use of sequencing, it has not replaced resistance phenotype assays because the presence of a gene does not ensure its expression. Therefore, we use resistance phenotype as a less exact way of detecting transmission of resistance within a hospital.

This article is organized as follows. In Section 2 we describe the data for our study: 592 multi-drug resistant isolates along with patient records from the community hospital. In Section 3 we examine the dependence of antibiotic resistance on factors other than time, by estimating a graphical model (Lauritzen 2004). We found no association between resistance and other patient factors (age, gender, infectious species, and sample type). This reduces each patient record to a pair consisting of the date of isolation and resistance phenotype. In Section 4 we introduce a mathematical model of transmission that assigns weights to pairs of such pairs. It uses the negative binomial distribution and is inspired by existing infection models. Parameters are learned from the medical literature. Section 5 is the heart of this paper. We introduce the NERD method, which tests whether the time course data are random or not, relative to our transmission model.

The p-values of this hypothesis test are computed for 16 antibiotics. For 13 antibiotics we found no significance, but for 3 antibiotics (Cefepime, Ceftazidime and Gentamicin) we did find significance. These findings are interpreted and analyzed further in Section 6. Section 7 offers a broader discussion of the meaning of our results, and whether it can say something about bacterial evolution in hospitals under drug pressure.

**2. Patient Data and Resistance Phenotypes**

A total of 592 Extended Spectrum β-lactamases (ESBLs) samples were collected from patients seen at Dignity Health Mercy Medical Center in Merced, California, between June 24, 2013 and January 23, 2016. ESBL strains are particularly interesting because they usually contain Class A β-lactamases that evolve very quickly and very specifically in response to clinical consumption of β-lactam antibiotics (Bush 2013). In a previous study (Mira, Crona et al. 2015), we identified β-lactam treatment plans that could reverse the evolution of ESBL resistant bacteria to penicillin and narrow spectrum β-lactam antibiotic. In this study we wanted to determine whether such a treatment plan makes sense in a hospital environment, or whether treatment plans should be used at a larger scale to manage antibiotic resistance.

**Table 1: Summary of Susceptibility Testing Results.**

|  | Susceptible | Intermediate | Resistant |
|---|---|---|---|
| Ampicillin | 0 | 0 | 470 |
| Ampicillin/Sulbactam | 85 | 113 | 284 |
| Piperacillin/Tazobactam | 429 | 46 | 30 |
| Cefazolin | 2 | 3 | 509 |
| Ceftazidime | 6 | 3 | 505 |
| Ceftriaxone | 6 | 0 | 508 |
| Cefepime | 7 | 2 | 505 |
| Ertapenem | 511 | 0 | 3 |
| Imipenem | 510 | 0 | 3 |
| Amikacin | 506 | 2 | 6 |
| Gentamicin | 343 | 2 | 169 |
| Tobramycin | 280 | 58 | 176 |
| Ciprofloxacin | 51 | 4 | 458 |
| Levofloxacin | 55 | 5 | 453 |
| Nitrofurantoin | 412 | 59 | 42 |
| Trimetprim/Sulfamethoxazole | 174 | 0 | 339 |

The samples were identified as ESBLs using Vitek 2 Version 06.01, an automated rapid detection system for pathogen identification and antibiotic sensitivity (Bobenchik, Deak et al. 2015). Following ESBL identification, the sensitivity to 16 antibiotics were also tested using broth microdilution minimum inhibitory concentration testing and the samples were categorized according to their susceptibility: Resistant (R), Intermediate (I), or Susceptible (S) based on the MIC Interpretation Guideline: CLSI M100-S26 (2015).

For each sample, we recorded 1) the date of sample isolation, 2) the age and the gender of the patient, 3) the species of the bacteria, 4) the tissue/source of the isolate, and 5) the susceptibility (R/I/S) to the following 16 antibiotics: Ampicillin, Ampicillin/Sulbactam, Piperacillin/Tazobactam, Cefazolin, Ceftazidime, Ceftriaxone, Cefepime, Ertapenem, Imipenem, Amikacin, Gentamicin, Tobramycin, Ciprofloxacin, Levofloxacin, Nitrofurantoin and Sulfamethoxazole/Trimetroprim. The isolate responses to each of the 16 antibiotics were organized into three possible categories: Susceptible (S), Intermediate (I) or Resistant (R). Table 1 gives the counts of these categories for each of the antibiotics.

Of the 592 records, 77 were incomplete or unreliable. We excluded these in our study. Among the 515 remaining records, most contained susceptibility testing for all 16 antibiotics. Some did not. This explains why the row sums in Table 1 are less than 515.

We end this section with a brief summary, aimed at a mathematician who is new to this subject. Our team worked with a hospital in Merced to obtain data. That data is a collection of about 500 de-identified patient records. A typical patient record looks like this:
( 05/17/2015, age 65, female, *E. coli*, urine, S,R,I,S,S,S,R,S,S,R,S,I,I,S,R,S )
The string of letters "S", "I" or "R" is the resistance phenotype with respect to the antibiotics. Here is the problem we are studying. For each of the 16 drugs separately, the title of this paper asks a question. Our goal is to find some answers, from these data alone.

### 3. Graphical Model

Our first step is to examine the dependence structure among the six discrete random variables: gender (male or female), age (by decades), tissue source of the sample (urine, blood, wound, or sputum), species of bacteria (*E. coli* or *K. pneumonia*), resistance phenotype (S, I, or R), and antibiotic. Graphical modeling is a statistical tool for studying dependence structures among several random variables (Lauritzen 2004). The question we seek to answer is whether any of the four variables gender, age, tissue, and species correlates with the resistance phenotype for a given antibiotic.

To this end, we organize the hospital data from Section 2 in a contingency table of format 2 x 10 x 4 x 2 x 3 x 16. We fit a graphical model to that table using the methods in the book *Graphical Models in R* (Højsgaard, Edwards et al. 2012). We use the algorithm described in Section 2.4 of this book. The algorithm searches through the space of all graphs and terminates when it has locally maximized the Akaike Information Criterion (AIC). It does not assign p-values to edges in the resulting graph.

The algorithm proceeds as follows. Starting from the full independence model (the graph with no edges), we compute the AIC at each stage of edge insertion. This was done using the function *forward* in the "gRim" package for the statistics software "R" (R Core Team 2013). This aims to ascertain the correct balance between numbers of edges (pa-rameters) and fit to the data. We also ran the algorithm that starts from the saturated model (the graph with all edges), and uses the *backward* function to compute the AIC of successive edge deletions. Both the forward algorithm and the backward algorithm terminated with the graph with six nodes and seven edges that is shown in Figure 1.

This preliminary analysis with graphical models suggests that Age, Tissue, Species and Gender are not correlated to the emergence of resistance. We therefore disregarded those variables in the subsequent analysis. For our study of the evolution of resistance, we used only the date of isolation, the antibiotic and the resistance information.

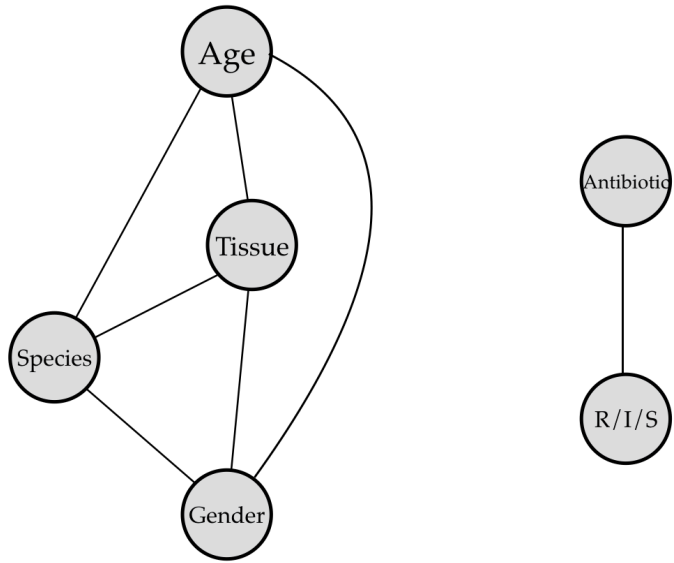

**Figure 1: Best fitting graphical model.** Each node is a discrete random variable. An edge between two nodes indicates statistical dependence. The first four nodes, (age, gender, tissues and species) have no edges to the last two nodes (antibiotics, R/I/S). The disconnectedness indicates statistical independence.

**4. Hospital Transmission Model**

Our study and hypothesis test rest on a probabilistic model we introduce for hospital-based transmission of antibiotic resistance between patients. We seek a model for patient interactions at the hospital, either direct or indirect, that have a causal effect on the resistance phenotype of the latter patient. In what follows, "transmission" is taken to mean "hospital-based transmission of antibiotic resistance from one patient to another".

The image we have in mind for our model is the following. There is a relationship between the resistance mechanisms carried by individual patient isolates and those carried by endemic bacteria at the hospital. When patients arrive, they are exposed to the endemic hospital bacteria. Resistance genes and/or the bacteria may then be transmitted to the patients, possibly after undergoing some evolution. When the resistance information of an individual patient isolate is measured, it gives an indication as to the resistance phenotype of the surrounding hospital bacteria. Our study uses the phenotypes of patient isolates as a proxy for understanding the resistance mechanisms of the hospi-tal bacteria. There are many complicated patterns of causality at play here. The model identifies those that are consistent with a change that takes place in the hospital.

*Single outbreak:*
Our model is based on the distribution of subsequent patient infections following a single bacterial clonal outbreak in a hospital. In such cases, whether the initial outbreak strain affects a patient depends on the time that has elapsed since the introduction of that strain. If the infectious strain of a latter patient is checked just hours after the outbreak strain's arrival, it is unlikely that there has been transmission. Likewise, transmission is unlikely if the latter patient visits the hospital several years after the original outbreak. Between these extreme time-points, likelihood of contamination is higher. We discretize the temporal information in units of "days". The random variable X is the number of days before a subsequent infection. We assume that X follows the negative binomial distribution. We do not restrict to patients affected directly by the original patient, but rather all subsequently affected patients. Note that often the negative binomial distribution is used to model the number of secondary patient infections, as in (Lloyd-Smith 2007). The justification for the use of the distribution in the present context is given below.

The negative binomial distribution is a discrete probability distribution on the set of non-negative integers k. It has two parameters p and r. The probability mass function is

$$P(x = k) = \binom{k + r - 1}{k} p^k (1 - p)^r.$$

Each trial can be either a success (with probability p) or a failure (with probability 1-p). The distribution models the number of successes before r failures. The parameter r is known as the over-dispersion parameter. The mean of the distribution is m = pr/(1-p).

The negative binomial distribution is applied to the random variable X, the number of days before a subsequent infection, according to the following rationale. Each day is a trial. Failure in a trial represents contact between a patient and endemic sources of resistance that day. The success probability p is the probability there is no contact. On the r'th instance of contact with a patient, transmission of endemic resistance occurs.

*Transmission of antibiotic resistance:*
We extend the scope of applicability of the negative binomial model to the set-up of our study. Starting at an initial patient isolate, the random variable X is the number of days before a transmission causes a latter isolate to have resistance phenotype caused by the first patient isolate. Transmission of antibiotic resistance may also involve a change in phenotype. General transmission cases are harder to detect than single outbreaks; our assumption is that the distribution is the same. The "single outbreak" case above is the specialization to where the first patient possesses a deadly bacterial strain, and we model the number of days until transmission of the strain to a latter patient occurs.

The two parameters in the negative binomial distribution affect the probabilities of the states of the random variable. The scale at which the process is being studied affects these probabilities. Transmission over larger scales than the hospital can occur after a larger number of days. At a global scale, transmission can sometimes be detected years later in another location (Medeiros 1997). The choice of parameters in our distribution specializes the set-up to hospital-based transmission.

We collected information from the medical literature to estimate the values of the parameters for our study. Table 2 gives observed statistical parameters for the occurrence of antibiotic resistance following a hospital outbreak. This is likely to depend on many factors about a hospital, some of which are included below. We also note that the variance far exceeds the mean, and as such the data are over-dispersed.

**Table 2: Information from clonal outbreaks and mean/variance computations**

| Outbreak strain | Size of hospital (beds) | Number of patients in outbreak | Average time from patient 0 (days) | Variance | Reference |
| --- | --- | --- | --- | --- | --- |
| *K. pneumoniae* | 880 | 127 | 439 | 37856 | (Arena, Giani et al. 2013) |
| *K. pneumoniae* | 1492 | 93 | 1027 | 49449 | (Giani, Arena et al. 2015) |
| *E. coli* | NA-community clinical lab | 69 | 340.6 | 30336 | (Pitout, Gregson et al. 2005) |
| *K. pneumoniae* | 243 | 17 | 108.6 | 2002 | (Snitkin, Zelazny et al. 2012) |
| *K. pneumoniae* | 81 | 7 | 122 | 4852 | (Kassis-Chikhani, Decre et al. 2006) |
| *K. pneumoniae* | 301 | 36 | 147 | 7886 | (Carrer, Lassel et al. 2009) |
| *E. coli* | 66 (NICU) | 21 | 103 | 4036 | (Garcia-Fernandez, Villa et al. 2012) |

We used the studies in Table 2 to estimate our parameters. The table indicates the importance of the size of the hospital in the mean and variance of the outbreak distribution. Dignity Health Mercy Medical Center has 186 beds. We fit our negative binomial parameters to the outbreak distribution of the hospital that was most similar in size, with 243 beds. This parameter fitting was done using the function *fitdistr* in the R package "MASS" (Venables and Ripley 2002). The parameters obtained were:

$$\text{Mean: } m = 115$$
$$\text{Over-dispersion: } r = 8.8$$
$$\text{Success probability: } p = m/(r+m) = 0.9289.$$

Our parameters say that with probability nearly 93%, there is no transmission on a given day. It could be that no patient arrived at the hospital that day, or that a patient did arrive but did not come into contact with the bacteria. The mean parameter of 115 days indicates that this is the mean length of time for a subsequent patient to be affected by an original patient. The effect may be indirect, and can proceed via other intermediate patient interactions. The parameter $r = 8.8$ indicates that, on average, on the 8.8'th instance of patient contact, there is transmission of antibiotic resistance.

## 5. Hypothesis Testing with the NERD Method

In this section we present the Nosocomial Evolution of Resistance Detector (NERD) method. This is a statistical hypothesis test, based on the model in Section 4. We apply it to the data described in Sections 2 and 3. In the context investigated here, the null hypothesis ($H_0$) and the alternative hypothesis ($H_1$) can be formulated as follows:

$H_0$: There is no evolution of antibiotic resistance at the hospital.
$H_1$: There is evolution of antibiotic resistance at the hospital, according to our model.

The NERD method works as follows. As described in Section 3, the data consist of antibiotic resistance information and temporal information, for each of our 16 antibiotics. The test is conducted for each antibiotic individually. For example, according to the second row of Table 1, the data for Ampicillin/Sulbactam is a sequence of 482 = 85 + 113 + 284 patient records. Each record is a pair. Concretely, the sequence looks like this:
    (0, R), (21, R), (21, R), (24, R), …. , (67, S), (67, I), (67, R), … , (963, R), (963, I)
The second and third pairs are patient records on day 21 with resistance phenotype "R".

We wish to assess the extent of non-randomness of this sequence, where departure from randomness is measured according to our model for hospital transmission. Detecting evolution according to that model minimizes the interference of antibiotic resistance evolution in the wider community. We perform this computation as follows.

For each of the nine combinations (SS, SI, SR, IS, II, IR, RS, RI, RR), we consider the pairs of records with these states. For example, the SR combinations are all pairs consisting of an earlier patient who is susceptible (S) to a particular antibiotic, and a later patient record with resistance (R) to that antibiotic. These include all possible causal transmissions from S to R in the data, but not all such pairs represent a causal transmission. Some are simply pairs SR that occur in the data by chance, or are due to evolution outside of the hospital. We assess the chance that a given SR pair was due to a true causal transmission, by assigning a weight to the occurrence of this pair. The weight is the probability mass function of the negative binomial distribution, with our estimated parameter values, evaluated at the elapsed time between the two records. The same is done for all nine combinations.

In summary, we compute a 3x3 table of transmission weights. Each weight is the sum of the probabilities $P(X = l - e)$ over relevant pairs of patient records where $e$ is the date of the earlier record and $l$ is the date of the later record. Here "relevant" means: for a fixed pair of states. In the example of the previous paragraph these are pairs ($e$,S) and ($l$,R).

We next normalize the 3x3 table of transmission weights by dividing each row by a constant so that it sums to one. The result is a table whose rows are probability distributions on the set (Alfredson and Korolik 2007). We refer to this as the table of empirical transmission probabilities. Its entries are the empirical probabilities of seeing an earlier patient with one resistance phenotype, and a latter patient with another resistance phenotype, with a change caused by the hospital environment, according to the model in Section 4.

This normalization in the previous step means that differences are concentrated on the final state of a pair, the state of the latter patient, rather than the state of the earlier pa-tient. We use it to focus our study on significance findings that pertain to the latter state, since these are an indication of changes occurring at the hospital.

At this stage of the NERD method, we have a table of empirical transmission probabilities. As an example, we show these probabilities for Ampicillin/Sulbactam in Table 3. The rows refer to the earlier patient and the columns to the latter patient.

**Table 3: Empirical probabilities for Ampicillin/Sulbactam**

|  | **Susceptible** | **Intermediate** | **Resistant** |
|---|---|---|---|
| **Susceptible** | 0.17782 | 0.23778 | 0.58439 |
| **Intermediate** | 0.16256 | 0.23457 | 0.60287 |
| **Resistant** | 0.17117 | 0.22806 | 0.60077 |

The null hypothesis assumption is that there is no departure from randomness in the data. This represents a lack of antibiotic resistance trend in the hospital, with time-delays that could be indicative of a hospital-based transmission. Under this assumption, we expect the three rows of the matrix to be identical. This is the mathematical meaning of the null hypothesis ($H_0$). The biological interpretation of ($H_0$) is that that future antibiotic resistance phenotype is not affected by the historical resistance phenotype, except via the average proportions of each resistance phenotype.

For the expected probabilities under ($H_0$), each row of the matrix is the proportion of patients of the corresponding resistance phenotype. This is obtained by dividing the number of counts S, I or R in the sequence by the length of the sequence. For Ampicillin/Sulbactam this length is 482, and we obtain the expected probabilities in Table 4.

**Table 4:**
**Expected probabilities for Ampicillin/Sulbactam under the null hypothesis**

|  | **Susceptible** | **Intermediate** | **Resistant** |
|---|---|---|---|
| **Susceptible** | 0.17635 | 0.23444 | 0.58921 |
| **Intermediate** | 0.17635 | 0.23444 | 0.58921 |
| **Resistant** | 0.17635 | 0.23444 | 0.58921 |

The observed values in Table 3 differ from the expected values in Table 4. To quantify this difference between the two 3x3 tables, we use the $\chi^2$ test statistic

$$\sum \frac{(observed - expected)^2}{expected}.$$

Here the sum is taken over all nine entries. For every 3x3 table whose rows are probability distributions on three states, the value of this test statistic is a nonnegative real number that measures the distance to the distribution expected under ($H_0$).

Our goal is to associate p-values to our data with respect to the $\chi^2$ test statistic. To do this, we use a permutation test to compare our test statistic with that of randomly gener-ated data. The p-value is the proportion of permutations whose $\chi^2$ test statistic is larger than the value computed in the data. In theory, the p-value could be obtained by considering all possible permutations of the observed resistance information, and finding the proportion of permutations that have larger $\chi^2$ value than our observed data.

For example, for the Ampicillin/Sulbactam data described above, the total number of possible permutations of the data sequence is given by the multinomial coefficient

$$\binom{482}{85, 113, 284} = \frac{482!}{85!\,113!\,284!} = 6 \times 10^{197}$$

This extremely large discrete problem can be approximated to good accuracy by generating 10,000 random permutations. We permuted the data accordingly, computed the $\chi^2$ test statistic, and compared it with the $\chi^2$ value for Table 3. The proportion of outliers was found to be 0.141. This is the p-value for the Ampicillin/Sulbactam data according to the NERD method. The analogous p-values for all 16 antibiotics are shown in Table 5.

In Table 5 we see that only three of the p-values are less than the commonly used threshold of 0.05, thus indicating significance for these three among the 16 antibiotics.

**Table 5: Hypothesis test p-value results.**

| Antibiotic | p-value |
| --- | --- |
| Ampicillin | All samples resistant |
| Ampicillin/Sulbactam | 0.141 |
| Piperacillin/Tazobactam | 0.426 |
| Cefazolin | 0.317 |
| Ceftazidime | **0.003*** |
| Ceftriaxone | 0.659 |
| Cefepime | **0.032** |
| Ertapenem | 0.395 |
| Imipenem | 0.403 |
| Amikacin | 0.565 |
| Gentamicin | **0.020** |
| Tobramycin | 0.116 |
| Ciprofloxacin | 0.096 |
| Levofloxacin | 0.123 |
| Nitrofurantoin | 0.192 |
| Trimethoprim/Sulfamethoxazole | 0.224 |

## 6. Interpretation

We used the NERD method to study each of the 16 antibiotics separately. For each antibiotic, we investigated the non-randomness of the data according to our hospital transmission model. For temporal information we used the patient record date (assumed to be approximately three days after the patient arrived at the hospital). The antibiotic resistance phenotype was measured on a three-point scale (R, I, S).

The results seen in Table 5 indicate that the hospital environment did not have a significant effect on the evolution of over ¾ of the antibiotics. However, the antibiotics Ceftazidime, Cefepime and Gentamicin have p-values below the 0.05. The star by the value for Ceftazidime indicates that this p-value remains significant after applying the Bonferroni correction to account for the 16 antibiotics tested. While the p-values of Cefepime and Gentamicin are not significant after this correction factor, their small p-values are still a point of interest in examining differences among our 16 antibiotics.

Since our study involved the two species of bacteria (*E. coli* and *K. pneumonia*) to determine whether the two species yielded different results we ran the method separately for each. We found that this had no effect on the significance of the p-values.

Having assessed the extent of departure from non-randomness under our model, we return to the data to examine the implications of this conclusion. The NERD method does not preferentially look for antibiotic resistance transmission of a particular kind. It does not preferentially detect changes from Susceptible to Resistant. Therefore, if there is a trend in antibiotic resistance over time, it remains to determine the directionality of this trend. We do so by introducing the notion of *relative cumulative resistance*.

The cumulative resistance at a given time point $t$ is the number of resistant isolates plus half the number of intermediate isolates recorded up to time $t$. We computed the actual curve from the data. The expected curve would be a straight line between zero resistance in June 2013, and the total cumulative resistance score in January 2016. The relative cumulative resistance is obtained by subtracting the expected cumulative resistance from the observed cumulative resistance.

If there are more susceptible patients near the start of the timeframe, and more resistant patients near the end, then the cumulative resistance curve is below the expected line, having a positive second derivative. Conversely, more resistant patients at the start of the time-window result in the cumulative resistance increasing more sharply at the start, before flattening off at the end of the timeframe. It will sit above the expected line and have a negative second derivative.

This analysis rests only on our data and does not depend upon any parameter choices. Since we consider all antibiotics across the same time frame, and we study qualitative comparisons between the trends, we do not worry about normalization of the y-axis: we do not wish to interpret particular values on the y-axis.

Figure 2 shows trend information for the three antibiotics with p-values less than 0.05. It demonstrates that the directionality of the resistance change does not match that of global trends.

From Figure 2, we conclude that trends in the frequencies of antibiotic resistant isolates in hospitals are not always increasing; they can indeed decrease as a result of evolution over time. The robust departure from randomness in our results indicates that the hospital environment can influence the frequencies of antibiotic resistant bacteria. The prescription of specific antibiotics may be relevant.

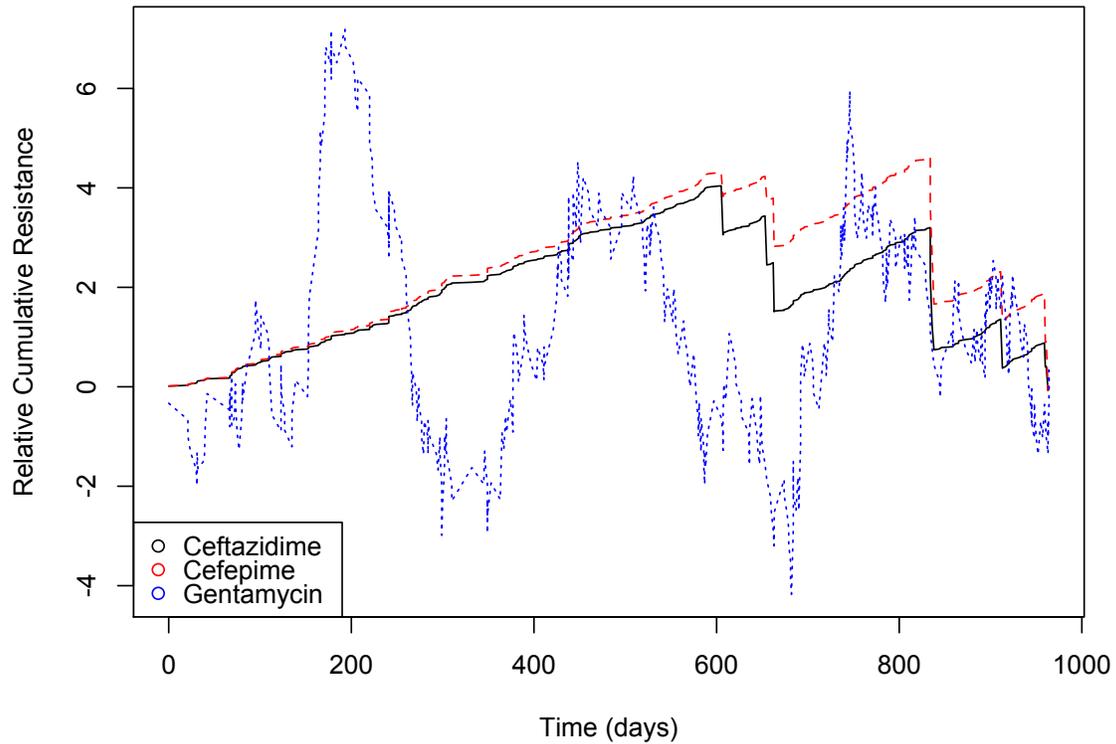

**Figure 2: Change in relative cumulative resistance of three antibiotics over time.** The graph plots the difference between the actual cumulative resistance, from our data, and the expected cumulative resistance. A value below 0 indicates an antibiotic with a trend from Susceptibility to Resistance over time; a value above 0 indicates a trend from Resistance to Susceptibility. Cefepime and Ceftazidime have a trend away from resistance. The trend for Gentamicin is neither positive nor negative overall. It follows an annual cycle, indicating seasonal variations in resistance phenotype.

## 7. Discussion

It is standard practice within hospitals to isolate infectious strains from patients and to determine their susceptibility to a panel of antibiotics relevant for treatment. This informs hospital staff of the best therapies available to individual patients. Additionally, the data enable hospitals to observe antibiotic resistance trends over time. Analysis of resistance trends is usually manual and consists of basic summary statistics of the bacterial populations aggregated over time. From such summaries, it is difficult to asses whether an antibiotic resistance trend is significant or not, and whether it is caused by immigration of bacteria into hospitals from the surrounding community.

We developed an automated quantitative method that produces a p-value to assess the significance of hospital-transmission trends while minimizing the effect of day-to-day variation in infectious strains brought in by patients. The method is based on the data that is often gathered by hospitals, and as such does not require them to expend resources gathering new data. The method incorporates a model of hospital transmission to determine whether resistance rates are a function of the hospital environment. We called this method the Nosocomial Evolution of Resistance Detector (NERD). It is available in the open-source R package "NERD", at https://github.com/seigal/NERD/.

We applied NERD to all 16 antibiotics in our study, and we drew conclusions independently for each of them. Since we did not specifically focus on those antibiotics that have low p-values, our p-values have not been adjusted to correct for multiple hypothesis testing. However, we note that one antibiotic remains significant after this correction factor, with significance implications. Future users of NERD who apply the method to search for antibiotics with evolving resistance will want to correct for multiply hypotheses using the Bonferroni correction.

For 13 of the 16 antibiotics, resistance phenotypes did not differ significantly from the antibiotic resistance phenotypes generated under the null hypothesis. This means that the hospital environment did not contribute to the evolution of antibiotic resistance more than outside factors. The histories of these antibiotics provide some context.

β-lactams (penicillins, cephalosporins, penicillin/inhibitors, carbapenems) were the most heavily used class of antibiotics for ~60 years, and they continue to be popular. Penicillins similar to amoxicillin are heavily used by the agricultural industry (Bailar and Travers 2002), and cephalosporin (Cefazolin and Ceftriaxone) consumption is far greater outside of the hospital than inside. Penicillin/inhibitor combinations are widely used in outpatient and hospital settings, but some such as Pipercillin/Tazobactam are mainly administered within hospitals. Consumption of carbapenems is more restricted than other β-lactams and mainly occurs within hospitals.

Fluoroquinolones (Ciprofloxacin and Levofloxacin) became the most heavily used antibiotics in the USA after the 2001 anthrax attacks (Navas 2002). Ciprofloxacin was the only antibiotic that had FDA approval for the treatment of anthrax. Their popularity has persisted, and their resistance has increased. They all share resistance mechanisms. Outpatient use of ciprofloxacin is at a much larger scale than hospital consumption.

Aminoglycosides (Tobramycin) received FDA approval in the 1970s and some have been used heavily in agriculture. Nitrofurantoin received FDA approval in 1953. Trimethoprim/Sulfamethoxazole received FDA approval in 1973. It can be used as an outpatient drug.

For the above antibiotics, large-scale consumption outside of the hospital probably limits the effect of hospital consumption. Resistance rates appear to be driven by bacterial populations outside the hospital. This accounts for their insignificant p-values.

Three of the antibiotics in our study were affected by the hospital environment. The evolution of Cefepime and Ceftazidime resistance is particularly striking. The similarities in relative cumulative resistance of Cefepime and Ceftazidime suggest that a single re-sistance gene is responsible for their resistance trends. There are not many resistance genes that simultaneously confer these two resistance phenotypes.

Cefepime is a cephalosporin type antibiotic that was introduced in 1994 for the treatment of moderate to severe infections such as pneumonia and urinary tract infections. It is only administered through injection and has no outpatient applications. Since Cefepime became widely used, resistant organisms have appeared. One of the first resistance genes found to hydrolyze Cefepime efficiently was CTX-M (Tzouvelekis, Tzelepi et al. 2000).

Numerous clinical strains of *E. coli* (Doi, Paterson et al. 2009, Mansouri, Ramazanzadeh et al. 2011) and *K. pneumonia* (Ambrose, Bhavnani et al. 2003, Wang, Hu et al. 2011) are resistant to Cefepime as a result of CTX-M expression. Additionally CTX-M has increased in frequency in clinical populations of bacteria and is now replacing other genes as the most commonly encountered. At the time of its first discovery, CTX-M did not confer Ceftazidime resistance, but subsequent mutations conferred this effect. In particular, the CTX-M-15 variant confers resistance to both Cefepime and Ceftazidime and is commonly detected in hospitals.

In a study for future publication (Doscher, Kim et al. 2016) we confirmed that presence of the CTX-M-15 gene is correlated with phenotype resistance to Cefepime and Ceftadizime. This demonstrates the strength of the NERD method in monitoring resistance genotypes. We performed whole genome sequencing for 48 isolates. For 39 of isolates (81.3%), resistance phenotype for ceftazidime and cefepime aligns with the status of CTX-M-15:
 26 isolates had the CTX-M-15 gene and were resistant or intermediate to both antibiotics, while 13 isolates had neither CTX-M-15 nor phenotype resistance. For 42 of isolates (87.5%), resistance aligns with status of any CTX-M variant: 32 isolates had a CTX-M gene and resistance to ceftazidime and cefepime, while 10 had neither. The correlation demonstrates that the NERD method can be used as a proxy for monitoring resistance genes.

Gentamicin is an aminoglycoside that has been available since the 1970s. Despite heavy use in agriculture, resistance rates for this antibiotic are moderate. It is also used in the treatment of urinary tract infections. Since the emergence of CTX-M resistance genes and Carbapenem resistant Enterobacteriaceae in urinary tract infections, non-β-lactam antibiotics have become necessary as primary treatment options for UTIs, and this may explain the significant Gentamicin resistance trend we observed.

Our results highlight that an individual hospital is an important but small piece of the overall resistance problem. Factors such as agricultural consumption of antibiotics, outpatient prescriptions and a high frequency of resistance genes in bacterial populations throughout the world also have a strong effect. Our results highlight the necessity of addressing antibiotic resistance at a larger scale. This may be at a community, regional, national, or global scale. Efforts at all levels are likely to help. For detection of trends within hospitals, from data that is routinely collected, the NERD method can be helpful.

**Acknowledgements:** We thank Dignity Health Mercy Medical Center and especially Robert Salcido, Glenn Bruss and Robert Streeter for providing clinical data and strains. We thank Joe Kileel for help organizing the data and defining the final data set. We are grateful to Lior Pachter for very helpful comments on an earlier version of this paper. This study was funded by CITRIS Seed Grant #: 2015-324. Anna Seigal and Bernd Sturmfels were supported by the National Science Foundation (DMS-1419018).